\documentclass[aps,prd,preprint,superscriptaddress,nofootinbib,amsmath,amssymb]{revtex4-1}

\usepackage{graphicx}
\usepackage{amsfonts}
\usepackage{multirow}

\begin{document}
\title{Three-dimensional black holes with conformally coupled scalar and gauge fields}
\author{Marcela C\'ardenas}
\email{cardenas@cecs.cl}
\affiliation{Centro de Estudios Cient\'{\i}ficos (CECs),  Casilla 1469, Valdivia, Chile.}
\affiliation{Departamento de F\'{\i}sica, Universidad de Concepci\'on, Casilla 160-C, Concepci\'on, Chile.}
 \author{Oscar Fuentealba}
\email{fuentealba@cecs.cl}
\affiliation{Centro de Estudios Cient\'{\i}ficos (CECs),   Casilla 1469, Valdivia, Chile.}
\affiliation{Departamento de F\'{\i}sica, Universidad de Concepci\'on, Casilla 160-C, Concepci\'on, Chile.}
 \author{Cristi\'an Mart\'{\i}nez}
\email{martinez@cecs.cl}
\affiliation{Centro de Estudios Cient\'{\i}ficos (CECs),   Casilla 1469, Valdivia, Chile.}
\affiliation{Universidad Andr\'es Bello, Av.~Rep\'ublica 440, Santiago, Chile.}

\begin{abstract}
We consider three-dimensional gravity with negative cosmological constant in the presence of a scalar and an Abelian gauge field. Both fields are conformally coupled to gravity, the scalar field through a nonminimal coupling with the curvature and the gauge field by means of a Lagrangian given by a power of the Maxwell one. A sixth-power self-interaction potential, which does not spoil conformal invariance is also included in the action. Using a circularly symmetric ansatz, we obtain black hole solutions dressed with the scalar and gauge fields,  which are regular on and outside the event horizon. These charged hairy black holes are asymptotically anti-de Sitter spacetimes. The mass and the electric charge are computed by using the Regge-Teitelboim Hamiltonian approach. If both leading and subleading terms of the asymptotic condition of the scalar field are present, a boundary condition that functionally relates them is required for determining the mass. Since the asymptotic form of the scalar field solution is defined by two integration constants, the boundary condition may or may not respect the asymptotic conformal symmetry.
An analysis of the temperature and entropy of these black holes is presented. The temperature is a monotonically increasing function of the horizon radius as expected for asymptotically anti-de Sitter black holes. However,  restrictions on the parameters describing the black holes are found by requiring the entropy to be positive, which, given the nonminimal coupling considered here, does not follow the area law.  Remarkably, the same conditions ensure that the conformally related solutions become black holes in the Einstein frame.
\end{abstract}
\maketitle

\section{Introduction}

In different contexts, a number of physical applications involving hairy black holes have emerged in the last years. For instance, asymptotically anti-de Sitter (AdS) black holes endowed with a scalar field have been related to superconductors by means of the gravity/gauge duality \cite{Horowitz:2010gk}. Additionally, in a totally different area,  efforts towards testing the no-hair theorem from astronomical observations have been recently developed \cite{Will:2007pp,Loeb:2013lfa}. The extensive literature about hairy black holes and the broad applications confirm their physical relevance.

Such as three-dimensional gravity has been a fruitful arena for quantum gravity, including the Ba\~{n}ados,  Teitelboim, and Zanelli (BTZ) black hole \cite{BTZ1,BTZ2}, it also has been very generous in providing exact black holes dressed with a scalar field. After the first examples \cite{Martinez:1996gn,Henneaux:2002wm} characterized by a scalar field which is regular everywhere,  other three-dimensional scalar hairy black holes have been reported with emphasis in the microscopic computation of their entropy \cite{CMT1,CMT2,Correa:2012rc} (see also \cite{Natsuume}), and as the result of an algorithm to determine all stationary circularly symmetric solutions \cite{Aparicio:2012yq}. The above results represent only a small part of the considerable attention that the three-dimensional scalar hairy black holes have received in recent years (see for instance \cite{Xu:2014uha} and references therein). 

In absence of a scalar field, the electrically charged BTZ black hole was introduced in \cite{BTZ1} and the rotating one with an electromagnetic field was presented in \cite{Martinez:1999qi}. In both cases the dynamics of the gauge field was defined by the usual Maxwell Lagrangian, and consequently, the gauge field exhibits a logarithmic dependence on the radial coordinate, as expected in $2+1$ dimensions. 

In this article we consider three-dimensional gravity with negative cosmological constant in the presence of a single real scalar field and an Abelian gauge field. This composite matter source is characterized by the fact that both fields are conformally coupled to gravity, in contrast with some recently proposed models \cite{Xu:2013nia,Mazharimousavi:2014vza}. The action for the scalar field contains, in addition to the kinetic term, an interacting term with the curvature and a sixth-power 
self-interaction potential. With these ingredients, this nonminimal action for the scalar field becomes conformal invariant. On the other hand, it is well known that the Maxwell action is invariant under conformal transformations of the metric only in four dimensions. This symmetry is recovered in any spacetime dimension $n$ if the Maxwell Lagrangian is raised to the $(n/4)^{\textrm{th}}$ power \cite{Hassaine:2007py}. Therefore, a Lagrangian of this form describes the Abelian gauge field considered in this work. Remarkably, this conformal invariant action for the gauge field may provide a Coulomb-like electric field in arbitrary dimensions.  

In the next section, we introduce the action and the corresponding field equations, which are solved using a circularly symmetric ansatz and the black hole solutions are identified. Since the solutions are given by simple expressions, the search for black holes is greatly simplified. 
Additionally, the geometrical and thermodynamic properties of them can be easily analyzed, and consequently, their physical meaning becomes clear.

The geometries asymptotically approach anti-de Sitter spacetime, and the scalar fields are regular on and outside of the corresponding horizons.  In Sec. \ref{mc} the mass and electric charge of the black holes are determined using the Regge-Teitelboim method \cite{Regge:1974zd}. Boundary conditions over the leading and subleading terms in the asymptotic form of the scalar field are required for obtaining the mass. Since the scalar field is defined for two independent integration constants, a wide class of boundary conditions are allowed, even those that spoil the asymptotic AdS invariance. Section \ref{thermo} is devoted to the thermodynamic analysis. The temperature, electric potential and entropy are determined. The entropy is not automatically a positive definite quantity in this nonminimal frame, and additional conditions must be imposed on the integration and self-interacting coupling constants in order to ensure a positive entropy. Finally, the last section contains some concluding remarks and prospects for future work.

\section{Black hole solutions} \label{bhs}

 We consider three-dimensional gravity with negative cosmological constant in presence of a scalar and an electromagnetic field, being both fields conformally coupled to gravity. The action is given by
\begin{equation} \label{action}
I[g_{\mu \nu}, \phi,  A_{\mu}]= \int d^3 x \sqrt{-g} \left[\frac{R+2l^{-2}}{2\kappa}-\frac{1}{2}g^{\mu \nu}\nabla_{\mu}\phi\nabla_{\nu}\phi-\frac{1}{16}R\phi^2-\lambda \phi^6+\sigma\left(-F^{\mu\nu}F_{\mu\nu}\right)^{3/4}\right],
\end{equation}
where $\kappa$ is the gravitational constant and $l$ is the AdS radius. Moreover,   $\lambda$ and $\sigma$ are the coupling constants of the self-interaction potential and the nonlinear electromagnetic term, respectively. 

The equations of motion are 
\begin{subequations}
\begin{eqnarray}
E_{\mu\nu}\equiv G_{\mu \nu}+\Lambda g_{\mu \nu}- \kappa\left( T_{\mu \nu}^{\phi}+T_{\mu \nu}^A\right) &=&0, \label{Eeq}\\
\Box\phi-\frac{1}{8}R\phi-6\lambda \phi^5 &=&0,\\
\partial_{\mu}\left(\sqrt{-g}\mathcal{F}^{-1/4}F^{\mu\nu}\right) \label{ME} &=&0,
\end{eqnarray}
\end{subequations}
with $F_{\mu\nu}=\partial_{\mu}A_{\nu}-\partial_{\nu}A_{\mu}$ and $\mathcal{F}=-F^{\mu\nu}F_{\mu\nu}$.
 
The energy-momentum tensor of the scalar field is given by
\begin{equation}\label{scalar-tensor}
T_{\mu\nu}^{\phi}=\partial_{\mu}\phi\partial_{\nu}\phi-\frac{1}{2}g_{\mu\nu}\partial^{\lambda}\phi\partial_{\lambda}\phi-\lambda g_{\mu\nu}\phi^6+\frac{1}{8}\left[g_{\mu\nu}\Box-\nabla_{\mu}\nabla_{\nu}+G_{\mu\nu}\right]\phi^2,
\end{equation}
and 
\begin{equation}\label{gauge-tensor}
T_{\mu\nu}^A=\sigma \left(3F_{\lambda\mu}F^{\lambda}_{\,\,\,\,\nu}\mathcal{F}^{-1/4}+g_{\mu\nu}\mathcal{F}^{3/4}\right)
\end{equation}
is the corresponding one for the nonlinear electromagnetic field.

It is worth noticing that the negative sign inside the nonlinear electromagnetic term in the action (\ref{action}) ensures that purely electric configurations remain real. Furthermore,  the coupling constant $\sigma$ is chosen to be positive\footnote{Without lost of generality, $\sigma$ is chosen to be $2^{1/4}$ just for simplifying  numerical factors.} in order to keep the energy density of the electromagnetic field --the $T^A _{\hat{0}\hat{0}}$ component of the energy-momentum tensor in the orthonormal frame-- positive for this class of configurations. 

Since the fields are conformally coupled, their corresponding stress tensors are traceless on-shell, so that Einstein's equations (\ref{Eeq}) imply  
\begin{equation} \label{R}
R=- 6 l^{-2}.
\end{equation} 
We will deal with asymptotically AdS spacetimes. In this context, potentials unbounded from below, for instance the case for $\lambda <0$ in the action (\ref{action}),  do not generate the sort of instabilities as in asymptotically flat spacetimes, provided the mass of the scalar field is bounded from below by the Breitenlohner-Freedman one $m_{BF}^2$ \cite{Breitenlohner:1982bm,Breitenlohner:1982jf,Mezincescu:1984ev}. In three dimensions, $m_{BF}^2=-l^{-2}$, and because of (\ref{R}), in our case the mass of the scalar field is $ 3/4 l^{-2}$, which satisfies the mentioned bound.

We look for static and circularly symmetric configurations described by the line element        
\begin{equation}\label{ds}
ds^2=-F(r)dt^2+F^{-1}(r)dr^2+r^2d\theta^2,
\end{equation}
a scalar field depending just on the radial coordinate, and a gauge field  of the form $A=A_t(r) dt$, which generates a purely radial electric field. The coordinates range as $-\infty < t < \infty, 0 \leq r < \infty, 0 \leq \theta < 2\pi$.

From the subtraction $E^t_t-E^r_r$ in (\ref{Eeq}), a second-order differential equation for the scalar field is obtained, whose integration yields
\begin{equation}
\phi(r)=\sqrt{\frac{b}{r+c}},
\end{equation}
where $b$ and $c$ are integration constants. Moreover, from the nonlinear Maxwell equation (\ref{ME}) the gauge field is easily obtained (modulo gauge transformations)
\begin{equation} \label{A}
A=-\frac{q}{r}dt.
\end{equation}
The constant $q$ is related with the electric charge as we will show below in Sec. \ref{mc}. Finally, the metric function $F(r)$ can be directly obtained from equation \eqref{R}, which gives 
\begin{equation}\label{F}
F(r)=\frac{r^2}{l^2}+a_1+\frac{a_2}{r},
\end{equation}
where $a_1$ and $a_2$ are integration constants. It is clear, from the line element (\ref{ds}) and the radial dependence of $F$  shown in (\ref{F}),  that these solutions are asymptotically anti-de Sitter spacetimes whose asymptotic behaviors match the well-known Brown-Henneaux conditions \cite{Brown-Henneaux}. However, as is discussed in Sec. \ref{mc}, boundary conditions on the matter fields could spoil the conformal invariance of the full configuration.

The case of vanishing scalar field ($b=0$) was studied in \cite{Cataldo:2000we}, and we will not consider it here. The remaining equations of motion give relations among the integration constants $a_1$, $a_2$, $b$, $c$ and $q$. Since we are interested in the case of nonvanishing scalar field we assume  $b \neq 0$ in these equations, which give rise to two different branches: i) $c=0$ and ii) $c\neq 0$. Furthermore, it is convenient to address the case without self-interaction potential ($\lambda=0$) in a separate section. 

\subsection{Case $c=0$: Black hole dressed with a stealth composite matter source} \label{c0}
The solution is determined by the metric function
\begin{equation}
F(r)=	\frac{r^{2}}{l^{2}} +24\lambda b^2,
\end{equation}
the scalar field
\begin{equation}
\phi(r)=\sqrt{\frac{b}{r}}, 
 \end{equation}
and the gauge field given by  (\ref{A}) with 
\begin{equation} \label{q1}
|q|^{3/2}=- \lambda b^3.
\end{equation}

The scalar field is real provided $b>0$. Moreover, in order to ensure a real $q$ it is necessary to fix the coupling constant $\lambda \leq 0$ as one can see from the r.h.s. of (\ref{q1}).  In this case, the spacetime corresponds to a black hole, whose horizon is located at  $r_{+}^2=- 24\lambda l^2 b^2$. It should be noticed that this black hole has the same metric as the static and uncharged BTZ black hole. However, it possesses a nonvanishing electric charge and is dressed with a conformal scalar field. This occurs because the total energy-momentum tensor vanishes, i.e., the scalar field and gauge field contributions cancel out. Therefore, the above solution can be considered as a stealth configuration \cite{AyonBeato:2004ig, AyonBeato:2005tu,Hassaine:2006gz,Maeda:2012tu,Ayon-Beato:2013bsa,Hassaine:2013cma} produced by two different matter sources. The metric is free of singularities and the matter fields diverge at the origin, $r=0$.

\subsection{Black holes in the general case $c, \lambda \neq 0$ } \label{cno0}
First, it is convenient to redefine the coupling constant as $\lambda=\kappa^2\alpha/(512 l^2)$, where now $\alpha$ plays the role of the coupling constant associated to the self-interaction potential. Additionally, we also define $b=8 a c /\kappa$, where $a$ is an integration constant.

In this way, the solution with a nonvanishing scalar field is given by the metric function
\begin{equation}\label{F2}
F(r)=\frac{r^{2}}{l^{2}}-\frac{(1-\alpha a^{2})}{l^{2}}\left(\frac{2c^{3}}{r}+3c^{2}\right),
\end{equation}
the scalar field
\begin{equation}\label{phi2}
\phi(r)=\sqrt{\frac{8}{\kappa}}\sqrt{\frac{ac}{r+c}},
\end{equation}
and the gauge field shown in (\ref{A}), with an integration constant $q$ satisfying
\begin{equation}\label{q}
|q|^{3/2}=-\frac{c^{3}(1-\alpha a^{2})(1-a)}{\kappa l^2}.
\end{equation}
This expression indicates that the gauge field vanishes for two particular values of $a$, which allow to rediscover previous \textit{uncharged} solutions. The case $a=1$  corresponds to the scalar hairy black hole found in \cite{Martinez:1996gn,Henneaux:2002wm}, and the case $a= 1/\sqrt{\alpha}$ is the massless hairy solution reported in \cite{Gegenberg:2003jr}.  In the case $a=0$, the scalar field vanishes and the extreme charged black hole in \cite{Cataldo:2000we} is recovered. Hereafter, we focus our attention in new black hole configurations with $q \neq 0$ and a nonvanishing scalar field.

The horizons are located at the positive roots of the cubic equation $F(r)=0$.  By replacing $r=c x$, this problem is reduced to solve
\begin{equation}\label{eqraices}
 x^{2}-(1-\alpha a^2)\left(\frac{2}{x}+3\right)=0.
\end{equation}
In the case $c>0$, we are interested in the positive roots of (\ref{eqraices}),  and for $c<0$ the relevant roots correspond to the negative ones. Since we are dealing with a cubic equation, it is possible to write down their exact roots $x_i$ in the following  form
\begin{equation} \label{roots}
x_{i}=z^{+}_{i}z^{-}_{i}\left(z^{+}_{i}+z^{-}_{i}\right),\qquad i=1,2,3
\end{equation}
with
\begin{equation}
z^{+}_{i}=\gamma_{i}(1+ \sqrt{\alpha a^2})^{1/3} \qquad \textrm{and} \qquad 
z^{-}_{i}=\bar{\gamma_{i}}(1- \sqrt{\alpha a^2})^{1/3}.
\end{equation}
Here $\gamma_{i}$ represent the roots of unity $\gamma_{i}^3=1$, and $\bar{\gamma_{i}}$ are their complex conjugates . These are 
\begin{equation}
\gamma_{1}=1,\qquad \gamma_{2}=-\left(\frac{1+i\sqrt{3}}{2}\right), \qquad \gamma_{3}=-\left(\frac{1-i\sqrt{3}}{2}\right).
\end{equation}
For $\alpha \leq 0$, $z^{-}_{i}=\bar{z}^{+}_{i}$  so that all the roots (\ref{roots}) are real. In the opposite  case $\alpha >0 $, we note that $x_1$ is always a real root of (\ref{eqraices}) and $x_2, x_3$ are complex. 

\begin{figure}[b] 
\centering 
\includegraphics[angle=0,width=0.40\textwidth]{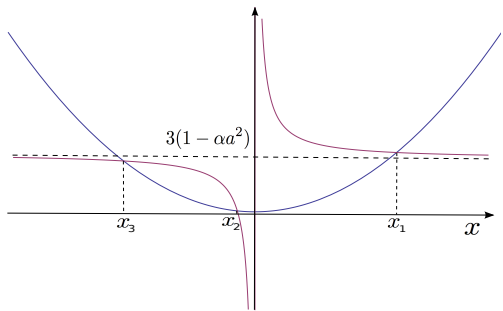}
\caption{The roots $x_1,x_2,x_3$ are shown for the case $(1-\alpha a^2)>0$. The root $x_1$ is positive and the roots  $x_2$ and $x_3$ depend on the sign of $\alpha$. If $\alpha<0$,  $x_2$ and $x_3$ are negative. Alternatively,  if $\alpha>0$,  $x_2$ and $x_3$ are both complex numbers, since the hyperbola does not intersect the parabola for $x<0$.}  
\label{fig:fig1}
\end{figure}

\begin{figure}
\centering
\includegraphics[angle=0,width=0.40\textwidth]{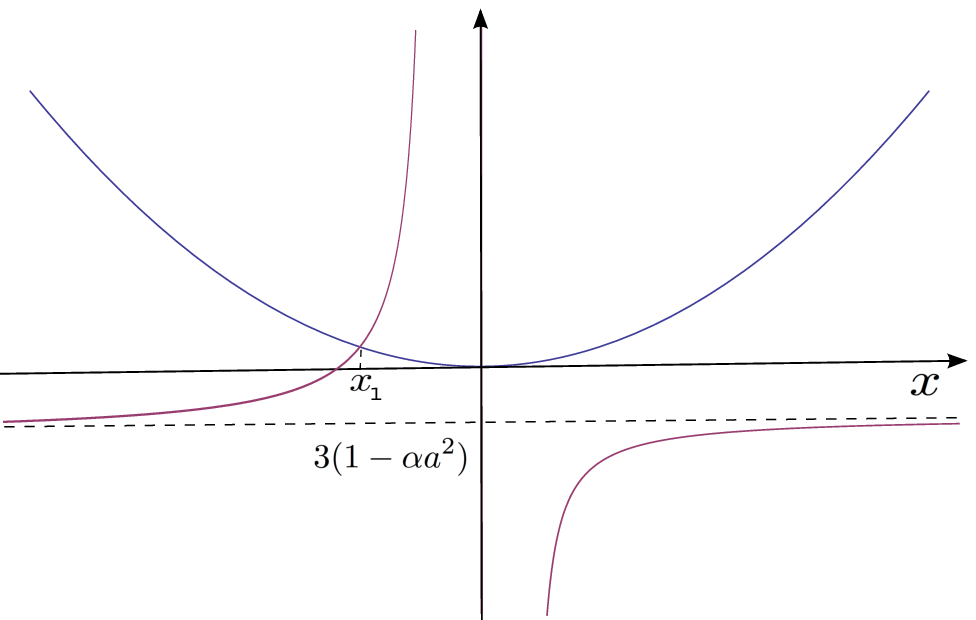} 
\caption{The roots $x_1,x_2,x_3$ are shown for the case $(1-\alpha a^2)<0$. The root  $x_1$ is negative and $x_2$ and $x_3$ are both complex roots.} 
\label{fig:fig2} 
\end{figure}

The qualitative behavior of the roots is illustrated in figures \ref{fig:fig1} and \ref{fig:fig2}. The real roots correspond to the intersection of a parabola with a hyperbola as is shown in (\ref{eqraices}). This can be described as follows:
\begin{itemize}
\item If $1-\alpha a^2>0$, the root $x_1$ is positive and  the nature of $x_2$ and $x_3$ depend on the sign of $\alpha$. If $\alpha<0$,  $x_2$ and $x_3$ are negative. On contrary,  if $\alpha>0$,  $x_2$ and $x_3$ are complex numbers (see Fig. \ref{fig:fig1}).    
\item If $1-\alpha a^2<0$, the root  $x_1$ is negative and $x_2$ and $x_3$ are complex roots (see Fig. \ref{fig:fig2}).
\end{itemize}
 
\vskip 0.5cm

After capturing the general properties of the roots of (\ref{eqraices}), we are in position to analyze the existence of black hole solutions. The analysis requires to study both signs of the integration constant $c$  as is shown below. Note that, for thermodynamic considerations explained in Sec. \ref{thermo}, the presence of a horizon is not enough to ensure physically sensible black hole solutions.

\subsubsection{Event horizon for $c>0$}

The previous analysis indicates that only for $1-\alpha a^2>0$ there is a positive root,  $x_1$.  Moreover, the condition $a >1$ appears  by demanding  positivity of the r.h.s  of (\ref{q}). The intersection of these two inequalities, $1-\alpha a^2>0$ and   $a >1$, implies that: 
\textbf{(A) }there is no restriction for any  $\alpha <0$, or 
\textbf{(B)} for a positive self-interacting coupling parameter $\alpha$, it is required to be bounded from above such that  $0<\alpha \leq 1$, in conjunction with a bounded integration  constant $1< a < 1/\sqrt{\alpha}$. 

Under the conditions  \textbf{(A)} or \textbf{(B)} there exists an event horizon located at $r_+= x_1 c$. Additionally, from the analytic expression of $x_1$ it is possible to determine bounds for the event horizon according to the sign of the self-coupling parameter. Under the conditions  \textbf{(A)} we have $r_+> 2c $,  while \textbf{(B)} provides the bounds $0<r_+<2c$.

Since $r$ and $c$ are positive, the scalar field is regular everywhere. The gauge field and metric are singular at the origin $r=0$, as one can read from (\ref{A}) and  from the Kretschmann invariant, $12 l^{-4}(1+2 c^6 (1-\alpha a^2)^2 r^{-6})$, respectively.

\subsubsection{Event horizon for $c<0$}

We are now interested in the negative roots of \eqref{eqraices}. First, the root $x_1 <0$ can be discarded since it requires the condition $1-\alpha a^2<0$ and also, from  (\ref{q}),  $a>1$. This last requirement is incompatible with the necessary  condition $a<0$ to ensure a real scalar field  \eqref{phi2}.  Therefore, $x_1$ does not produce an event horizon. Second, it is possible to consider the roots $x_2$ and $x_3$, which become negative real numbers provided $1-\alpha a^2>0$ and $\alpha<0$ (conditions labeled by \textbf{(C)}).  From the definitions of $x_2$ and $x_3$ one can to extract the following properties: $ 2/3 < | x_2| <1$ and $ | x_3|>1$. Then, since $|x_3| > |x_2| $ the event horizon is located at $r_+=  x_3 c $, provided conditions \textbf{(C)} are satisfied. The root $x_2$ gives rise an inner horizon. Since we are considering $\alpha$ and $ a \neq 0$, the root  $ x_2$ cannot  equal $x_3$, then an extreme black hole does not occur. Due to the inequality $r_+ > -c$, the singularity of the scalar field at $r=-c$ is hidden by the event horizon $r_+$. As in the previous case, the metric and gauge field are singular only at the origin.

\subsection{ Black hole in absence of self-interaction potential ($\lambda=0$)}  \label{l0}

A particularly simple solution is obtained in absence of self-interaction potential. The metric function $F(r)$ reduces to
\begin{equation}\label{F3}
F(r)=\frac{(r+c)^{2}(r-2c)}{r l^{2}},
\end{equation}
and the gauge and scalar fields are given by (\ref{A}) and  (\ref{phi2}), respectively. Now, the constant $q$ satisfies
\begin{equation}\label{q3}
|q|^{3/2}=\frac{c^{3}(a-1)}{\kappa l^2}.
\end{equation}
Although it is possible to consider $c<0$, the double zero of $F(r)$ at $r=-c$ is not suitable to be promoted to event horizon because the scalar field  (\ref{phi2}) is singular there. We adopt a conservative point of view saying that the singularity of the scalar field prevents the existence of an extreme black hole. Thus, we consider only the simple root $r=2c$, which becomes an event horizon for $c>0$. The condition $a>1$ arises from (\ref{q3}).  As in the previous case with $c > 0$, the gauge and metric fields are singular at the origin, and the scalar field is regular everywhere.

\section{Mass and electric charge} \label{mc}
The aim of this section is to determine the conserved charges of the black holes introduced above. For this goal we consider the Hamiltonian Regge-Teitelboim method \cite{Regge:1974zd}.  In this approach the charges $Q[\xi,\xi^{A}]$ are the surface terms added to the Hamiltonian generator in order to ensure well-defined functional derivatives. The bulk piece of the canonical generator 
\begin{equation}\label{H}
H[\xi,\xi^{A}]=\int dx^2\left(\xi^{\perp} \mathcal{H}_{\perp}+\xi^{i} \mathcal{H}_{i} +\xi^{A} \mathcal{G}\right)+Q[\xi,\xi^{A}],
\end{equation}
is a linear combination of the constrains $\mathcal{H}_{\perp}$ and $\mathcal{H}_i$, where $i$ denotes the two spatial dimensions, and  $\mathcal{G}$ is the Gauss constraint associated to the Abelian gauge field. The charge corresponds to the canonical generator for vanishing constraints. The vector $\xi=(\xi^{\perp}, \xi^{i})$ represents the asymptotic symmetries of the spacetime, and $\xi^{A}$ is the parameter associated to the Abelian gauge  symmetry.

For the class of solutions we are dealing with, it is sufficient to consider a minisuperspace of circularly symmetric configurations defined by the line element
\begin{equation}
ds^2=-\left(N^{\perp}(r)\right)^2 dt^2+F(r)^{-1}dr^2+r^2d\theta^2,
\end{equation}
a scalar field $\phi(r)$ and a gauge field $A=A_t(r) dt$. In this case, the only nontrivial canonical momentum is that corresponding to the gauge field $\mathcal{E}(r)$, which is given by
\begin{equation}
\mathcal{E}(r)=3 r \left( \frac{F(r)}{\left(N^{\perp}(r)\right)^2}\right)^{1/4}| F_{t r}|^{1/2}  \, \mbox{\textrm{sign$(F_{t r})$}}.
\end{equation}
Since all the canonical momenta associated to the gravitational field and the scalar field vanish,  the constraint $\mathcal{H}_{i}$ is identically zero, $\mathcal{H}_{\perp}$ takes the form
\begin{equation}
\mathcal{H}_{\perp}=\frac{1}{\sqrt{F}}\left(\frac{F' }{2\kappa}\left(1 -\frac{\kappa\phi^{2}}{8}\right)+\frac{rF}{4}(\phi'^{2}-\phi\phi'')-\frac{\phi\phi'}{8}(rF'+2F)-\frac{r}{l^2\kappa}+r \lambda \phi^{6}+\frac{\mathcal{|E|}^{3}}{27 r^{2}}\right),
\end{equation}
and the Gauss constraint reduces to $\mathcal{G} =-\partial_r\mathcal{E}$.

The variation of surface term $\delta Q$ is chosen so that $\delta H =0$ on the vanishing constraints. In this case,  the boundary is a circle $S^{1}$ of infinite radius. Integrating over the angular coordinate, we obtain
\begin{eqnarray} \label{dQ}
\delta Q(\xi^{\perp},\xi^{A})&=&\left[\frac{\pi  \xi^{\perp}\left(-8+\kappa  \phi  \left(\phi +2 r \phi' \right)\right)}{8 \kappa \sqrt{F}}\delta F -\frac{1}{2}\pi r\sqrt{F} \left(\phi \left(\xi^{\perp}\right)'+3\xi^{\perp} \phi' \right)\delta\phi \right. \\ \nonumber
&+& \left. \frac{1}{2} \pi  r \sqrt{F} \xi^{\perp}\phi \delta \phi'+2 \pi  \xi^{A} \delta \mathcal{E} \right]_{r\rightarrow \infty}.
\end{eqnarray}
The integration of $\delta Q$ requires to choose suitable asymptotic conditions for all fields. These conditions should allow for the asymptotic behavior of the exact solutions found in the previous section. These conditions, specified up to the order that contributes to the charge, are given by 
\begin{eqnarray}
F(r) & =& \frac{r^2}{l^2}+F_{0}+\mathcal{O}\left(\frac{1}{r}\right),\\
\phi(r) & =& \frac{\phi_0}{r^{1/2}}+\frac{\phi_1}{ r^{3/2}}+\mathcal{O}\left(\frac{1}{r^{5/2}}\right), \label{pa}\\
\mathcal{E} (r) &=&\mathcal{E} _{0} +\mathcal{O}\left(\frac{1}{r}\right),\\
\xi^{\perp}(r) & =& \frac{r}{l}\xi_0+\mathcal{O}\left(\frac{1}{r}\right),\\
\xi^{A}(r) &=& \xi^{A}_{0} +\mathcal{O}\left(\frac{1}{r}\right),
\end{eqnarray}
where the quantities labeled with subscripts $0$ and $1$ are arbitrary constants.
Under these asymptotic conditions the variation (\ref{dQ}) reduces to 
\begin{equation}
\delta Q = \xi_0\left( -\frac{\pi \delta F_0 }{\kappa} +\frac{\pi}{2 l^2}\left(3 \phi_1 \delta \phi_0-  \phi_0 \delta \phi_1\right) \right)+ 2 \pi \xi^{A}_{0} \delta \mathcal{E}_0. \label{dQ1}
\end{equation}
The mass $M$ is the conserved charge associated to time translation symmetry, parametrized here by $\xi_0$, and the electric charge $Q_{\textrm{e}}$ is that coming from the U(1) gauge invariance, represented  by the gauge parameter $\xi^{A}_{0}$. From (\ref{dQ1}) we can read directly
\begin{eqnarray}
\delta M & =& -\frac{\pi \delta F_0 }{\kappa} +\frac{\pi}{2 l^2}\left(3 \phi_1 \delta \phi_0-  \phi_0 \delta \phi_1\right), \label{dM1}\\
\delta Q_{\textrm{e}} &=& - 2 \pi \delta \mathcal{E}_0.  \label{dqe1}
 \end{eqnarray}
The minus sign in (\ref{dqe1}) comes from the sign difference between the electric field density and  the canonical momentum of the gauge field.
The electric charge can be immediately integrated, and is given by the leading term of the canonical momentum of the gauge field:
\begin{equation} \label{ec}
 Q_{\textrm{e}} = - 2 \pi  \mathcal{E}_0.
\end{equation}
It is clear that the second term in (\ref{dM1}), which takes into account the contribution of the scalar field to the mass,  provided $\phi_0 \neq 0$ and $\phi_1 \neq 0$, needs a boundary condition for integrating it, i. e., a functional relation $\phi_1= \phi_1(\phi_0)$. In simple words, the mass is determined after imposing boundary conditions, and is given by\footnote{Two arbitrary additive constants (but fixed, i. e. without variation) appear in the integration of (\ref{dM1}) and (\ref{dqe1}). They will be set to zero in order that the massless BTZ has a vanishing mass, and in absence of the gauge field, the solution be electrically uncharged.}
\begin{equation} \label{mass}
 M = -\frac{\pi  F_0 }{\kappa}+\frac{\pi}{2l^2}\int \left( 3\phi_1 -\phi_0\frac{d \phi_1}{d \phi_0}\right) d \phi_0.
\end{equation}
Apart from the boundary conditions $\phi_0 = 0$ or $\phi_1 = 0$, there is only one additional case which also leads a vanishing contribution from scalar field to the mass: the functional relation 
\begin{equation}
\phi_1=\gamma \phi_0^3,
\end{equation}
where $\gamma $ is a constant without variation. These three boundary conditions share a same feature: they do not spoil the conformal invariance of a scalar field approaching to infinity in the form (\ref{pa}), as pointed out in \cite{Henneaux:2002wm} (for a recent related discussion in four dimensions see \cite{Anabalon:2014fla}). Any other functional relation $\phi_1= \phi_1(\phi_0)$, in the way of Designer Gravity \cite{Hertog:2004ns},  breaks the conformal invariance of the scalar field and consequently, the asymptotic AdS symmetry of the whole configuration.

We can now compute the mass and electric charge for the black holes found in Section \ref{bhs}. The first case is the black hole with stealth matter described in section \ref{c0}. In this case, 
$ F_0= 24 \lambda b^2, \phi_0= \sqrt{b}, \phi_1=0$ and $\mathcal{E}_0=3\lambda^{1/3} b  \, \mbox{\textrm{sign$(q)$}}$. Then, evaluating \eqref{mass} and \eqref{ec}, the corresponding mass and electric charge are
\begin{equation}
M=  -\frac{ 24\pi  \lambda b^2 }{\kappa}, \quad \mbox{and} \quad Q_{\textrm{e}}=6\pi (-\lambda)^{1/3} b  \, \mbox{\textrm{sign$(q)$}},
\end{equation}
respectively.

For the black holes found in section \ref{cno0}, $\phi_0, \phi_1 \neq 0$ and a boundary condition is required in order to determine the mass. For instance, the boundary condition  
\begin{equation}\label{bc}
\phi_1=\gamma \phi_0^n,
\end{equation}
where $\gamma, n\neq -1$ are constants without variation, yields a mass
\begin{equation}
M=-\frac{\pi  F_0 }{\kappa}+\frac{\pi\gamma(3-n)}{2l^2(n+1)}\phi_0^{n+1}.
\end{equation}
Then, using the asymptotic values for this class of black holes,
\begin{equation}
F_0=-\frac{3c^2(1-\alpha a^2)}{l^2},\qquad \phi_0=\sqrt{\frac{8 ac}{\kappa}},\qquad \phi_1=-\sqrt{\frac{2ac^3}{\kappa}},
\end{equation}
the mass and the electric charge can be written as
\begin{eqnarray} 
M&=&\frac{3\pi c^2(1-\alpha a^2)}{\kappa l^2}+\frac{\pi \gamma(3-n)}{2l^2(n+1)}\left(\frac{8ac}{\kappa}\right)^{\frac{n+1}{2}}, \label{2}\\
Q_{\textrm{e}}&=&  6\pi |q|^{1/2} \, \mbox{\textrm{sign$(q)$}},
\label{3}\end{eqnarray}
where $q$ is given in Eq. (\ref{q}). Note that the boundary condition (\ref{bc}) fixes a relation between the integration constants $a$ and $c$. Finally, the limit $\alpha \rightarrow 0$ in (\ref{2}) and  (\ref{3}) gives the mass and electric charge of the black hole without self-interaction potential described in Section \ref{l0}.

\section{Thermodynamics} \label{thermo}

This section is devoted to study thermodynamic properties of the charged hairy black holes shown in Sec. \ref{bhs}. The conjugate variables associated to the conserved charges, mass and electric charge,  are the temperature and the electric potential, respectively.  The temperature can be obtained by means of the surface gravity $\kappa_{\textrm{H}}$ 
\begin{equation}\label{Temperature}
T=\frac{\kappa_{\textrm{H}}}{2\pi},
\end{equation}
which is given by $\kappa_{\textrm{H}}^2=-1/2\nabla^{\mu}\chi_{\nu}\nabla_{\mu}\chi^{\nu}$ with $\chi^{\mu}=\left(1,0,0\right)$. Additionally, the electric potential is defined as
\begin{equation} \label{voltage}
\Phi:=A_0(r_+)-A_0(\infty)=-\frac{q}{ r_+}.
\end{equation}

The entropy can be found using the modified Bekenstein-Hawking area formula,
\begin{equation}\label{entropy}
S=\Omega(r_+)\frac{4\pi^2 r_+}{\kappa},
\end{equation}
where the factor $\Omega(r_+)=1-\kappa \phi(r_{+})^2/8$ comes from the nonminimally coupling term  
in the action  \cite{Visser:1993nu,Ashtekar:2003jh}. Since this factor is
not positive definite, the entropy could become negative. In order to avoid such a non-well-behaved thermodynamic  situation, solutions in which $\Omega(r_+)$ is negative must be discarded as black holes.  For this reason, it is necessary to impose additional constrains on the integration constants and $\alpha$ as discussed in detail below.

We start examining the validity of the first law for the black holes introduced in Sec.\ref{bhs}. Using the variation of the global charges \eqref{dM1} and \eqref{dqe1}, and the expressions for the temperature (\ref{Temperature}), entropy (\ref{entropy}) and electric potential (\ref{voltage}), evaluated on each particular black hole,  it is possible to prove that the first law of black hole thermodynamics
\begin{equation}
\delta M=T\delta S+ \Phi \delta Q
\end{equation}
holds in all the cases. It can be seen as follows. In the general case $c\neq 0$  the expressions for each member of the above equation are given by
\begin{eqnarray}
\delta M &=&-\frac{2\pi c^2}{l^2\kappa}\left(1+3\alpha a\right)\delta a+\frac{6\pi c}{l^2\kappa}\left(1-\alpha a^2\right)\delta c\\
\delta Q_e &=& -\frac{2^{5/6} \pi \textrm{sign$(q)$} \sigma ^{2/3} c \left(3 \alpha  a^2-2 \alpha  a-1\right) \delta a}{\kappa  l^2 (1-a)^{2/3} \left(1-\alpha  a^2\right)^{2/3}}\nonumber\\& &-\frac{3\ 2^{5/6} \pi  \textrm{sign$(q)$} \sigma ^{2/3} \left(1-a\right)^{1/3} \left(1-\alpha  a^2\right)^{1/3} \delta c}{\kappa  l^2}\\
\delta S &=& -\frac{4\pi^2 x\delta a}{\kappa (1+x)}-\frac{4\pi^2(-1+a-x)x\delta c}{\kappa (1+x)}+\frac{4\pi^2 c(-a+(1+x)^2)\delta x}{\kappa (1+x)^2}.
\end{eqnarray}
After applying repeatedly the following identities which comes from $F(r_+)=0$,
\begin{equation}
x^3=(1-\alpha a^2)(2+3x),\qquad \delta x=-\frac{2\alpha a (2+3x)\delta a}{3(-1+\alpha a^2+x^2)},
\end{equation}
it is possible to show that the first law is satisfied. Note that this property holds regardless a relation between $a$ and $c$, i. e., the first law is satisfied for any boundary condition. For $c=0$ the check of the first law is straightforward, we have in this case
\begin{eqnarray}
\delta M &=&-\frac{48\pi \lambda b \delta b}{\kappa},\\
\delta Q_e &=& 6\pi (-\lambda)^{1/3}\textrm{sign$(q)$}\delta b,\\
\delta S &=& \frac{8l\sqrt{6|\lambda|}\delta b}{\kappa}.
\end{eqnarray}
Now, we analyze the thermodynamic behavior of the black hole solutions according to the different values of $c$.
In the case $c= 0$, discussed in Sec. \ref{c0}, there exists an event horizon only if the coupling constant $\lambda$ is negative. Then, the temperature, electric potential and entropy are 
\begin{equation}\label{thermo1}
T=\frac{r_+}{2 \pi l^2 },\qquad
\Phi=-\frac{\textrm{sign$(q)$}r_+}{24(-\lambda)^{1/3}l^2},\qquad 
S=\left(1-\frac{\kappa}{8 l \sqrt{24(-\lambda)}} \right)\frac{4\pi^2 r_+}{\kappa},
\end{equation}
respectively. We can see that these quantities are linear functions of $r_+=2\sqrt{6|\lambda|}lb$. However, the entropy is positive only if $\sqrt{-\lambda}<\kappa/(8l\sqrt{24})$. Thus, this physical  requirement on the entropy yields an upper bound for the coupling constant $\lambda$.

\begin{figure}[h] 
\centering 
\includegraphics[angle=0,width=0.40\textwidth]{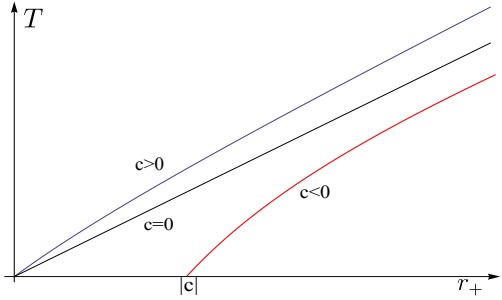} 
\caption{The  behavior of the temperature $T$ as a function of the horizon radius $r_+$ is shown. For all possible values of the integration constant $c$,  the temperature is a monotonically increasing function of $r_+$.  For large values of $r_+$, $T$ approaches a linear function of $r_+$ with a slope  $1/(2\pi l^2)$, which matches the straight line describing the case $c=0$.  Note that for a given temperature, three possible black hole configurations of different sizes can exist.} \label{fig:figtemp}
\end{figure}

For the case  $c\neq 0$, studied in Sec. \ref{cno0},  the expressions for temperature,  electric potential, and  entropy computed from  (\ref{Temperature}), (\ref{entropy}), and (\ref{voltage}) are
\begin{eqnarray}\label{thermo2}
& &T=\frac{3r_+}{2\pi l^2}\left(\frac{r_++c}{3r_++2c}\right),\;
\Phi=-\frac{ \, \mbox{\textrm{sign$(q)$}}c^{2}(1-\alpha a^{2})^{2/3} (1-a)^{2/3}}{(\kappa l^2)^{2/3}r_+},\; \\ & & S=\left(1-\frac{ac}{r_{+}+c}\right)\frac{4\pi^{2}r_{+}}{\kappa}, \label{ent2}
\end{eqnarray}
respectively.

First, we analyze the temperature behavior. Since $r_++c >0$, the temperature is a positive,  monotonically increasing function of $r_+$ as shown in Fig. \ref{fig:figtemp}. For large values of $r_+$, the temperature approaches a linear function of $r_+$ with the same slope, $1/(2\pi l^2)$, as for that in the case $c=0$, which coincides with the temperature of the static BTZ black hole.

Now, we focus the attention on the entropy (\ref{ent2}) for $c\neq0$.  As mentioned, the entropy derived from the action (\ref{action}) is not a positive definite quantity. Then, the conditions that guarantee black holes with positive entropy must be determined. This is summarized in Table \ref{tab:tabla}.

\begin{table}[htb!]
\begin{center}
\begin{tabular}{|c|c|c|c|}
\cline{2-4}
\multicolumn{1}{c|}{}& Horizon conditions  &$\alpha$& $a$\\
\hline
\multirow{3}{*}{$c>0$}&$0<\alpha \leq 1$  &\multirow{2}{*}{$0<\alpha \leq 1$ }& \multirow{2}{*}{$ a<a^*$}\\
&$1<a\leq 1/\sqrt{\alpha}$&&\\
\cline{2-4}
&$\alpha \leq 0$&$-1/3<\alpha \leq 0$&$ a<a^*$\\
\cline{3-4}
&$a>1$&$-\infty<\alpha \leq -1/3$&  No condition\\
\hline
\multirow{2}{*}{$c<0$}& $\alpha<0$   &$-3<\alpha<-1/3$&$ a<a^*$\\
\cline{3-4}
&$a<0$& $-\infty<\alpha \leq -3$&No condition\\
\hline
\end{tabular}
\caption{The table shows the conditions that ensure a positive entropy for $c\neq 0$. In the first column the existence conditions for black holes are displayed. The second and third columns exhibit the range of the coupling parameter $\alpha$ and the integration constant $a$ for which  the entropy is positive. Here  $a^*= (3+\alpha)/(1+3\alpha)$. We observe that a large negative $\alpha$ ensure a well-defined entropy without extra conditions. As is explained in the text, these conditions also ensure that the mapped solutions, using (\ref{conft}), correspond to black holes in the Einstein frame.} \label{tab:tabla}
\end{center}
\end{table}

A conformal transformation maps the action (\ref{action}) to the Einstein frame (EF), where the scalar field is minimally coupled to gravity. 
The entropy in the Einstein frame follows the Bekenstein-Hawking area law, and hence is a positive definite quantity. Naively, one may think that negative entropy configurations, now mapped into the Einstein frame could have a positive entropy. Remarkably, as shown in \cite{Martinez:2005di}, for a similar class of solutions in four dimensions, they are not mapped into black holes but naked singularities in the new frame. The mechanism acts as follows. In three dimensions the conformal transformation is given by
\begin{equation}
g^{\textrm{EF}}_{\mu\nu}=\Omega(\phi)^2 g_{\mu\nu}=\left(  1-\frac{\kappa}{8}\phi^{2}\right)^2 g_{\mu\nu}\label{conft}
\end{equation}
First, we note the hypersurfaces where the conformal factor $\Omega(\phi)$ vanishes are mapped into curvature singularities of the corresponding image in the EF. For the solutions presented in this work,  $\Omega(\phi)$ is a monotonously increasing function of $r$ approaching $1$ for a large $r$, but it is not a positive definite function. For configurations where  $\Omega(\phi(r_+))\leq 0$, the conformal factor necessarily vanishes in a hypersurface located at $r_0 \geq r_+$ generating a naked singularity in the EF. On the contrary, for those configurations with  $\Omega(\phi(r_+))> 0$, the curvature singularity occurs at $r_0 < r_+$. Only in the latter case black holes in the conformal frame are mapped into black holes in the EF.  In consequence, since $\Omega(\phi(r_+))> 0$ is the same condition for ensuring a positive entropy, only black holes having a well-defined entropy in the conformal frame generate black holes in the EF. Therefore one concludes that the conditions for the black
holes parameters in the conformal frame, shown in Table \ref{tab:tabla} ensuring the entropy to
be positive,  exactly coincide with the ones that guarantee cosmic censorship in Einstein frame. 

\section{Discussion} \label{dis}

We have obtained exact, circularly symmetric, three-dimensional black holes, which are regular on and outside their event horizons, endowed with conformally coupled scalar and gauge fields. We remark that all these interactions were fixed ab initio in the action. The black holes are described by means of very simple expressions, even in the presence of a self-interaction potential compatible with the conformal invariance. For this reason, their geometries and thermodynamic properties can be easily explored, and consequently, the physical meaning of them becomes clear.

In general, the integration of the field equations provides two arbitrary constants which parametrize the solutions in conjunction with the self-interaction coupling constant. The black holes can be classified in three groups. The first group, discussed in Sec. \ref{c0}, includes those with a stealth composite matter source,  where the contributions of both fields to the energy-momentum tensor cancel out. The case in which the three parameters do not vanish defines the second group treated in Sec. \ref{cno0}. Here two black holes appear, one with a single horizon, and another one having an inner horizon, which cannot become extreme, keeping a nontrivial scalar field. The third group is defined by the absence of the self-interaction potential (Sec. \ref{l0}). This class contains the electrically charged version of the black hole found in \cite{Martinez:1996gn}.  Additionally, an extreme black hole emerges if the condition of regularity for the fields at the horizon is removed.

It is worth noting that the asymptotic behavior of the metrics satisfies the Brown-Henneaux asymptotic conditions even in the case of a nontrivial scalar and gauge fields. This means that these are asymptotically AdS spacetimes. However, the \emph{entire} configuration is endowed with the asymptotic AdS invariance only if the scalar field allows it.

The conserved charges, mass and the electric charge, were determined under the Regge-Teitelboim approach. It was found that boundary conditions on the leading and subleading terms of the asymptotic form of the scalar field are necessary in order to obtain the mass. This fact is in accordance with the physical statement which says that the mass is well defined after boundary conditions are imposed.  

Remarkably, the scalar fields presented in sections \ref{cno0} and \ref{l0} have an asymptotic behavior allowing to analyze a wide class of boundary conditions, even including those that break the asymptotic AdS symmetry. This is possible because the scalar field contains two independent integration constants unlike other exact solutions as far we know, which are defined with only one integration constant and hence no other boundary condition is required.  These black holes could be considered in the context of the so-called Designer Gravity theories \cite{Hertog:2004ns}, in which general boundary conditions were numerically studied.  However, since the black holes shown here are exact solutions, these could be very useful for those models.

The temperature of the black holes is a monotonically increasing function of the horizon radius $r_+$, which approaches the linear one for large $r_+$ as it happens in general for the AdS black holes. On the other hand, the factor appearing in the modified entropy area law is not necessarily positive definite. Hence the positiveness of the entropy requires extra conditions on the integration constants and the coupling parameter $\alpha$ as shown in Table \ref{tab:tabla}. We note that for a large enough negative coupling constant the entropy is positive without other conditions apart of those necessary for the existence of black holes. Since in the Einstein frame the entropy is a positive definite quantity, one may think that negative entropy configurations could have a well-defined thermodynamic description in that frame as well. However, this class of solutions is mapped to naked singularities in the Einstein frame. It is worth pointing out the exact correspondence between the positiveness of the entropy in the conformal frame and the cosmic censorship principle in the Einstein frame. The black holes in the Einstein frame, and their geometrical and thermodynamic properties deserve further attention and they are interesting enough as to be considered in a future work. Finally, in three dimensions adding angular momentum is not a difficult task, and it would be interesting to study the spinning versions of the black holes introduced here.

\acknowledgments

We thank Hern\'an A. Gonz\'alez,  Javier Matulich, David Tempo, Ricardo Troncoso and Jorge Zanelli for helpful discussions. M. C. and O. F. thank Conicyt for financial support. This work has been partially funded by the  Fondecyt grants 1121031 and 1130658.
The Centro de Estudios Cient\'{\i}ficos (CECs) is funded by the Chilean Government
through the Centers of Excellence Base Financing Program of Conicyt.

\end{document}